\documentclass[prb,twocolumn,superscriptaddress]{revtex4}
\usepackage{graphicx}
\usepackage{amsmath}
\usepackage{amsfonts}
\usepackage{amssymb,mathrsfs}

\newcommand{\bs}[1]{\boldsymbol{#1}}

\begin{document}

\title{Counterflows in viscous electron-hole fluid}

\author{P.S. Alekseev}
\affiliation{A.F. Ioffe Physico-Technical Institute, 194021 St. Petersburg, Russia}
\author{A.P. Dmitriev}
\affiliation{A.F. Ioffe Physico-Technical Institute, 194021 St. Petersburg, Russia}
\author{I.V. Gornyi}
\affiliation{Institut f\"ur Nanotechnologie,  Karlsruhe Institute of Technology,
76021 Karlsruhe, Germany}
\affiliation{\mbox{Institut f\"ur Theorie der kondensierten Materie,  Karlsruhe Institute of
Technology, 76128 Karlsruhe, Germany}}
\affiliation{A.F. Ioffe Physico-Technical Institute, 194021 St. Petersburg, Russia}
\affiliation{L.D. Landau Institute for Theoretical Physics, Kosygina street 2, 119334 Moscow, Russia}
\author{V.Yu. Kachorovskii}
\affiliation{A.F. Ioffe Physico-Technical Institute, 194021 St. Petersburg, Russia}
\affiliation{L.D. Landau Institute for Theoretical Physics, Kosygina street 2, 119334 Moscow, Russia}
\affiliation{Institut f\"ur Nanotechnologie, Karlsruhe Institute of Technology,
76021 Karlsruhe, Germany}
\author{B.N. Narozhny}
\affiliation{\mbox{Institut f\"ur Theorie der kondensierten Materie, Karlsruhe Institute of
Technology, 76128 Karlsruhe, Germany}}
\affiliation{National Research Nuclear University MEPhI (Moscow Engineering Physics Institute),
  115409 Moscow, Russia}
\author{M. Titov}
\affiliation{Radboud University Nijmegen, Institute for Molecules and Materials, NL-6525 AJ
Nijmegen, The Netherlands}
\affiliation{ITMO University, 197101 St. Petersburg, Russia}

\affiliation{}

\date{\today}

\begin{abstract}
   In ultra-pure conductors, collective motion of charge carriers at
   relatively high temperatures may become hydrodynamic such that
   electronic transport may be described similarly to a viscous
   flow. In confined geometries (e.g., in ultra-high quality
   nanostructures), the resulting flow is Poiseuille-like. When
   subjected to a strong external magnetic field, the electric current
   in semimetals is pushed out of the bulk of the sample towards the
   edges. Moreover, we show that the interplay between viscosity and
   fast recombination leads to the appearance of counterflows. The edge
   currents possess a non-trivial spatial profile and consist of two
   stripe-like regions: the outer stripe carrying most of the current
   in the direction of the external electric field and the inner
   stripe with the counterflow.
\end{abstract}

\maketitle

Recently, signatures of the hydrodynamic behavior of charge carriers
have been observed in graphene \cite{exg1,exg2,exg3}, palladium
cobaltate \cite{exp}, and the Weyl semimetal WP$_2$ \cite{exw}. This
phenomenon occurs in the intermediate temperature regime, where the
typical length scale of electron-electron interaction, $\ell_{ee}$, is
much shorter than any other relevant scale in the problem including
those characterizing scattering off potential disorder and
electron-phonon scattering, ${\ell_{ee}\ll\ell_{\rm
    dis},\ell_{ph}}$. In this case, the independent particle
approximation is violated, the motion of charge carriers becomes
collective, and transport properties of the system are determined by
interaction \cite{us1,luk}.

Viscous electronic fluids exhibit unusual transport properties
\cite{us1,luk}, such as superballistic transport \cite{exg3,mor,lev},
nonlocal resistivity \cite{exg1,fal}, and negative magnetoresistance
\cite{exw,gul,pal,moore,als}. The latter effect may also occur in
two-component systems (e.g., semimetals or narrow-band semiconductors)
near the charge neutrality point \cite{usp}. In such systems, response
of the charge carriers to the external magnetic field is non-universal
depending on the interplay between inelastic scattering processes
and sample geometry.

In the hydrodynamic regime, electronic transport can be described with
the help of the linearized hydrodynamic theory \cite{pal,moore,als,usp}
generalizing the standard Navier-Stokes equation \cite{dau6}. The
parameters of the theory, including the shear viscosity coefficient,
$\eta_{xx}$, and quasiparticle recombination time, $\tau_R$, can be
derived, at least in principle, from the kinetic equation approach
(for a particular case of graphene, see Ref.~\onlinecite{hyd}). Due to
the above two processes, the electric current density in a
finite-sized sample is nonuniform. In long samples (where the length
is much larger than the width, ${L\gg{W}}$), viscous effects tend to
form a Poiseuille-like flow. The actual profile of the current density
depends on the ratio of the typical length scale describing the
viscous effects, the so-called Gurzhi length \cite{usp}, $\ell_G(B)$,
and the sample width, $W$. In the limit where the Gurzhi length
exceeds the width, ${\ell_G\gg{W}}$, the current density profile is
parabolic, similarly to the standard viscous flow \cite{dau6,poi}. In
the opposite case, the current density profile resembles the catenary
curve \cite{usp}, where significant inhomogeneities are localized at
the sample edges. In both cases, the electric charge is being
transmitted mostly through the bulk, avoiding the edges (this effect
is the physical origin of the superballistic transport found in
Refs.~\onlinecite{exg3,lev}).

\begin{figure}
\centerline{\includegraphics[width=0.8\columnwidth]{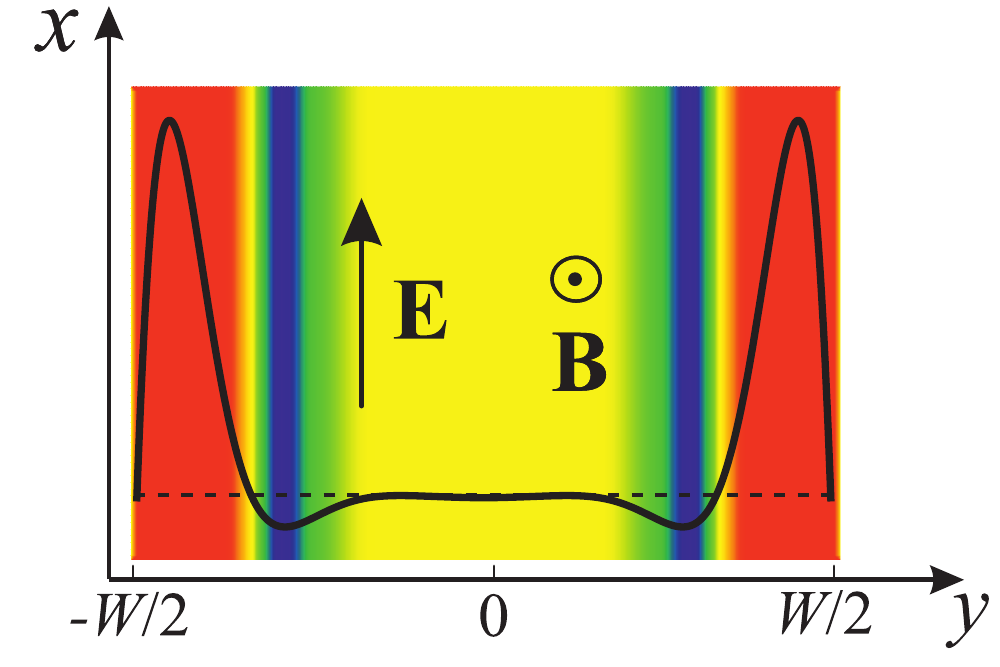}}
\caption{Schematic plot of the inhomogeneous electric current density
  in the regime of fast recombination and strong enough magnetic
  field. The color map emphasises the positive (i.e., parallel to the
  external electric field) current at the edge (red) contrasted to the
  negative (i.e., opposite to the external electric field) current in
  the intermediate stripe (blue). The black curve illustrates the
  magnitude of the current density at a given point along the
  sample. The dashed line indicates the zero value of the current.}
\label{fig1:j0}
\end{figure}

Two-component systems may possess an additional inelastic scattering
process: the electron-hole recombination. This process is known to
create a boundary layer \cite{us2,us3,meg} characterized by linear
magnetotransport \cite{gus,mol,vas}. In general, the recombination
boundary layer coexists with the above viscous boundary layer. In the
hydrodynamic regime, the typical time scale describing the
recombination processes is much longer than the electron-electron
relaxation time, ${\tau_R\gg\tau_{ee}}$. Since the latter defines the
Gurzhi length in the absence of the magnetic field, this can be recast
in the relation of the corresponding length scales,
${\ell_R\gg\ell_G(0)}$. Both length scales decrease with the
applied magnetic field. The decrease of the Gurzhi length follows from
the field dependence of the shear viscosity \cite{pal,moore} and is
governed by the electron-electron scattering. In contrast, the
effective length scale associated with the recombination processes
follows from the solution of the hydrodynamic equations
\cite{usp,us2,us3} and is governed by the dominant elastic scattering
process. In the previous paper \cite{usp}, we have considered the
limit of weak (or slow) recombination, where $\tau_R$ is much longer
than the elastic mean free path and hence the recombination length is
much longer than the Gurzhi length for an arbitrary magnetic field,
${\ell_R\gg\ell_G}$. In this limit, the system exhibits
unconventional transport properties. For typical parameter values, the
magnetoresistance of a long sample is a nonmonotonic function of the
field: it is negative in weak fields and then becomes positive and
linear in strong fields.

In this paper, we consider the opposite limit of relatively fast
recombination, such that the recombination time, $\tau_R$, is much
smaller than the elastic mean free path. We show that in this case the
electric current density is strongly inhomogeneous and, in contrast to
the standard Poiseuille flow, is mostly concentrated at the sample
edges. The structure of the edge currents is most peculiar and
consists of two regions, see Fig.~\ref{fig1:j0}. While the wider,
outer region carries the large current in the direction of the applied
electric field, the current in the narrower inner region flows in the
{\it opposite} direction. The latter counterpropagating current is
much smaller than the former, such that the total current is parallel
to the electric field. However, this system exhibits a most curious
example where a local current density is directed opposite to the
external electric field.

Although we are focusing on a specific model of a compensated
semimetal, the phenomenon of the counterflow is more general. Similar
effects have been suggested in the context of the {\it ac} transport
\cite{moe,aac}. Counterflows in a steady state, thermoelectric flow in
a single-component electron fluid (e.g. in doped graphene or a
semiconductor) will be discussed in a subsequent publication \cite{usf}.

\section{Hydrodynamics of compensated semimetals}

The hydrodynamic model of a two-component conductor with electron-hole
recombination was discussed in Ref.~\onlinecite{usp}. Here we repeat
the main points for completeness.

Recombination processes violate the particle number conservation
for each individual constituent of the system. As a result, the
continuity equations have the form
\begin{equation}
\label{ceq}
\frac{\partial \delta n_{\alpha}}{\partial t}
+ \bs{\nabla}\cdot\bs{j}_{\alpha}
 = -\frac{ \delta n_{e}+\delta n_{h}}{2\tau_R},
\end{equation}
where $\alpha=e,h$ distinguishes the type of carriers, $\delta
n_\alpha$ are the deviations of the carrier densities from their
equilibrium values $n_\alpha^{(0)}$, $\bs{j}_{\alpha}$ are the carrier
currents, and $\tau_R$ is the electron-hole recombination time.

In the hydrodynamic regime, charge transport can be described by the
generalized Navier-Stokes equation. Within linear response, the
equations for the two constituent of the system have the form
\cite{usp}
\begin{eqnarray}
\label{eq0}
&&
\frac{\partial\bs{j}_\alpha}{\partial t}
+
\frac{\langle v^2 \rangle}{2}\bs{\nabla}\delta n_{\alpha}
-
\frac{e_\alpha n_\alpha^{(0)}}{m} \bs{E}
-
\omega_\alpha \left[ \bs{j}_\alpha \times \bs{e}_z \right]
=
\\
&&
\nonumber\\
&&
\qquad\qquad\qquad\qquad\qquad
=
-\frac{\bs{j}_\alpha}{\tau}
-\frac{\bs{j}_\alpha - \bs{j}_{\alpha'}}{2\tau_{eh}}
+
\eta_{xx} \Delta \bs{j}_\alpha.
\nonumber
\end{eqnarray}
Here we consider the orthogonal magnetic field,
${\bs{B}\!=\!B\bs{e}_z}$; the electron and hole charges are
${e_h\!=\!e>0}$, ${e_e\!=\!-e}$, and the cyclotron frequencies are
${\omega_\alpha\!=\!e_\alpha{B}/(mc)\!=\!\omega_ce_\alpha/e}$; the
index $\alpha'$ denotes the constituent other than $\alpha$:
${\alpha'\!=\!e}$ for ${\alpha\!=\!h}$ and vice versa; $\tau_{eh}$ is
the momentum relaxation time due to electron-hole scattering; and the
averaging (for the parabolic spectrum with the constant density of
states $\nu_0$) is defined as \cite{usg}
\[
\langle\dots\rangle \!=\!-\!\!\int\!\!d\epsilon \, 
\frac{\partial f^{(0)}(\epsilon)}{\partial\epsilon} (\dots),
\]
where $f^{(0)}(\epsilon)$ is the Fermi distribution function. The
choice of the parabolic bands simplifies the algebra, but is not
essential; all qualitative features of our results remain valid for an
arbitrary spectrum (respecting the rotational invariance \cite{jul}).

The field-dependent shear viscosity is given by
\cite{pal,st}
\begin{equation}
\label{eta_xx}
\eta_{xx} \!=\! \eta_0/(1\!+\!4\omega_c^2 \tau_{ee}^2),
\end{equation}
where $\eta_0$ is the shear viscosity in the absence of the magnetic
field 
\begin{equation}
\label{eta0}
\eta_0 \!=\! \langle v^4\rangle\tau_{ee}/(4\langle v^2\rangle)
\sim \langle v^2 \rangle \tau_{ee}.
\end{equation}
The off-diagonal (or Hall) viscosity is neglected in Eq.~(\ref{eq0})
since the corresponding contribution is much smaller than the Lorentz
terms, see Ref.~\onlinecite{usp} for details.

The hydrodynamic theory is justified if the electron-electron
scattering time $\tau_{ee}$ is the shortest time scale in the problem
(including the ``ballistic'' time defined by the sample width)
\begin{equation}
\label{cond}
\tau_{ee}\ll \tau, \tau_R, \tau_{eh}, \tau_W,
\qquad
\tau_W\sim W/\sqrt{\langle v^2\rangle}.
\end{equation}
In this case, the equations (\ref{eq0}) describe the two (electron and
hole) fluids that are weakly coupled by electron-hole scattering
\cite{mor,us2,us3,svi}. Unlike the single-component fluid considered
in Ref.~\onlinecite{pal}, these two fluids cannot be considered as
incompressible. However, under the assumption (\ref{cond})
electron-hole recombination dominates the viscous compressibility
(related to bulk viscosity) allowing us to drop the latter from
Eqs.~(\ref{eq0}).

In this paper we restrict our consideration to charge neutrality,
${n_e=n_h}$. Introducing the total quasiparticle density,
${\rho\!=\!n_e\!+\!n_h}$ and the linear combinations of the two
currents, ${\bs{P}\!=\!\bs{j}_e\!+\!\bs{j}_h}$ and
${\bs{j}\!=\!\bs{j}_h\!-\!\bs{j}_e}$, we re-write the hydrodynamic
theory (\ref{ceq}) and (\ref{eq0}) as
\begin{subequations}
\label{heqs}
\begin{equation}
\label{ceq1}
\frac{\partial\delta\rho}{\partial t} + \bs{\nabla}\cdot\bs{P}
 = -\frac{\delta\rho}{\tau_R},
\qquad
\bs{\nabla}\cdot\bs{j}=0,
\end{equation}
\begin{equation}
\label{eqp}
\frac{\partial\bs{P}}{\partial t}
+
\frac{\langle v^2 \rangle}{2}\bs{\nabla}\delta\rho
-
\omega_c \left[ \bs{j} \times \bs{e}_z \right]
=
-\frac{\bs{P}}{\tau}
+
\eta_{xx} \Delta\bs{P},
\end{equation}
\begin{eqnarray}
\label{eqj}
\frac{\partial\bs{j}}{\partial t}
-
\frac{e\rho^{(0)}}{m} \bs{E}
-
\omega_c \left[ \bs{P} \times \bs{e}_z \right]
=
-\frac{\bs{j}}{\tau}
-\frac{\bs{j}}{\tau_{eh}}
+
\eta_{xx} \Delta \bs{j}.
\end{eqnarray}
\end{subequations}
Here ${\rho\!=\!\rho^{(0)}\!\!+\!\delta\rho}$ with
${\rho^{(0)}\!=\!n_e^{(0)}\!\!+\!n_h^{(0)}}$.

Finally, in the long sample geometry, ${L\gg{W}}$, all physical
quantities are functions of the transverse coordinate $y$ only. At
charge neutrality, the total electric field is equal to the applied
field, ${\bs{E}\!=\!(E,0)}$. Requiring that no current flows out of
the sides of the sample, ${j_y(\pm{W}/2)\!=\!P_y(\pm{W}/2)\!=\!0}$, we
find that the electric current is directed along the strip,
${\bs{J}\!=\!e\bs{j}\!=\!e(j(y),0)}$, while the total quasiparticle
flow, ${\bs{P}\!=\!(0,P(y))}$, is orthogonal. As a result, we arrive
at the steady state equations \cite{usp}
\begin{subequations}
\label{heqs2}
\begin{equation}
\label{ceq2}
P' = -\delta\rho/\tau_R,
\end{equation}
\begin{eqnarray}
\label{eqp2}
&&
\langle v^2 \rangle\delta\rho'/2
+
\omega_c j
=
-P/\tau
+
\eta_{xx} P''
\end{eqnarray}
\begin{eqnarray}
\label{eqj2}
&&
- e\rho^{(0)} E/m
- \omega_c P
=
-j/\tau
-j/\tau_{eh}
+
\eta_{xx} j''
\end{eqnarray}
\end{subequations}

Excluding the quasiparticle density, $\delta\rho$, we find the two
coupled differential equations describing the electric current density
and the lateral neutral quasiparticle flow
\begin{subequations}
\label{heqs4}
\begin{equation}
\label{eqj4}
\ell_G^2(B) j''-j+\sigma_0E+\omega_c\tau_*P=0,
\end{equation}
\begin{equation}
\label{eqp4}
\ell_R^2 P''-P-\omega_c\tau j=0,
\end{equation}
where field-dependent Gurzhi length 
\begin{equation}
\label{lg}
\ell_G(B) = \sqrt{\eta_{xx}\tau_*}
= \sqrt{\eta_0\tau_*/[1\!+\!(2\omega_c \tau_{ee})^2]},
\end{equation}
characterizes the viscous effects, while
\begin{equation}
\label{lr}
\ell_R\!=\!\sqrt{(\eta_{xx}\!+\!\langle v^2\rangle\tau_R/2)\tau}
\!\approx\!\sqrt{\langle v^2\rangle\tau_R\tau/2},
\end{equation}
\end{subequations}
describes the recombination [the latter equality follows from
  Eq.~(\ref{cond})]. The quantity $\sigma_0$ has the meaning of the
zero-field conductivity of an infinite sample given by 
\begin{equation}
\label{s0}
\sigma_0 = e\rho^{(0)}\tau_*/m, \qquad \tau_* = \tau\tau_{eh}/(\tau+\tau_{eh}).
\end{equation}
The mean free time $\tau_*$ reflects the combined effect of the
disorder scattering and mutual electron-hole friction.

The equations (\ref{heqs4}) allow for a formal solution [assuming the
standard no-slip boundary conditions ${j(\pm{W}/2)=0}$]
\begin{subequations}
\label{r2}
\begin{equation}
\label{rjp2}
\begin{pmatrix}
j \cr
P
\end{pmatrix}
\!=\!
\left[ 1 \!-\!
\cosh (\widehat{M}^{\frac{1}{2}}y)\!\left[\cosh(\widehat{M}^{\frac{1}{2}}W/2)\right]^{-1}\right]\!\!
\begin{pmatrix}
j_0 \cr
-\omega_c\tau j_0
\end{pmatrix}\!,
\end{equation}
where the matrix $\widehat{M}$ is given by
\begin{equation}
\label{m}
\widehat{M} =
\begin{pmatrix}
\ell_G^{-2}(B) & -\omega_c\tau_*\ell_G^{-2}(B) \cr
\omega_c\tau\ell_R^{-2} & \ell_R^{-2}
\end{pmatrix}.
\end{equation}
The spatial variation of the currents is governed by the eigenvalues
of the matrix (\ref{m})
\begin{eqnarray}
\label{lambda}
&&
\lambda_\pm \!=\!
\left[\ell_G^{-2}(B)\!+\!\ell_R^{-2}\right]/2 \pm\!
\\
&&
\nonumber\\
&&
\qquad
\pm 
\sqrt{\!\left[\ell_G^{-2}(B)\!-\!\ell_R^{-2}\right]^2/4\!
-\!\ell_G^{-2}(B)\ell_R^{-2}\omega_c^2\tau\tau_*}
\!.
\nonumber
\end{eqnarray}
Using the eigenvalues (\ref{lambda}), we express the current
densities, $j(y)$ and $P(y)$, as
\begin{widetext}
\begin{equation}
  \label{rj2}
  j \!=\! \frac{j_0}{\lambda_+\!-\!\lambda_-}\!
  \left[\!\left(1\!-\!\frac{\cosh \sqrt{\lambda_+}y}{\cosh \sqrt{\lambda_+}W/2}\right)\!
    \!\left[\ell_G^{-2}(B)(1\!+\!\omega_c^2\tau\tau_*)\!-\!\lambda_-\right]
    \!-\!\left(1\!-\!\frac{\cosh \sqrt{\lambda_-}y}{\cosh \sqrt{\lambda_-}W/2}\right)
    \!\left[\ell_G^{-2}(B)(1\!+\!\omega_c^2\tau\tau_*)\!-\!\lambda_+\right]\!
    \right]\!,
\end{equation}
\begin{equation}
  \label{rp2}
  P \!=- \frac{\omega_c\tau j_0}{\lambda_+\!-\!\lambda_-}\!
  \left[\lambda_+\!\left(1\!-\!\frac{\cosh \sqrt{\lambda_-}y}{\cosh \sqrt{\lambda_-}W/2}\right)
      -\lambda_-\!\left(1\!-\!\frac{\cosh \sqrt{\lambda_+}y}{\cosh \sqrt{\lambda_+}W/2}\right)\!
       \right]\!,
\end{equation}
\end{widetext}

\noindent
where
\begin{equation}
\label{j0}
j_0 = \sigma_0 E/(1+\omega_c^2\tau\tau_*),
\end{equation}
\end{subequations}
is the uniform current density in an infinite sample.

\begin{figure*}[t!]
\centerline{\includegraphics[width=0.4\textwidth]{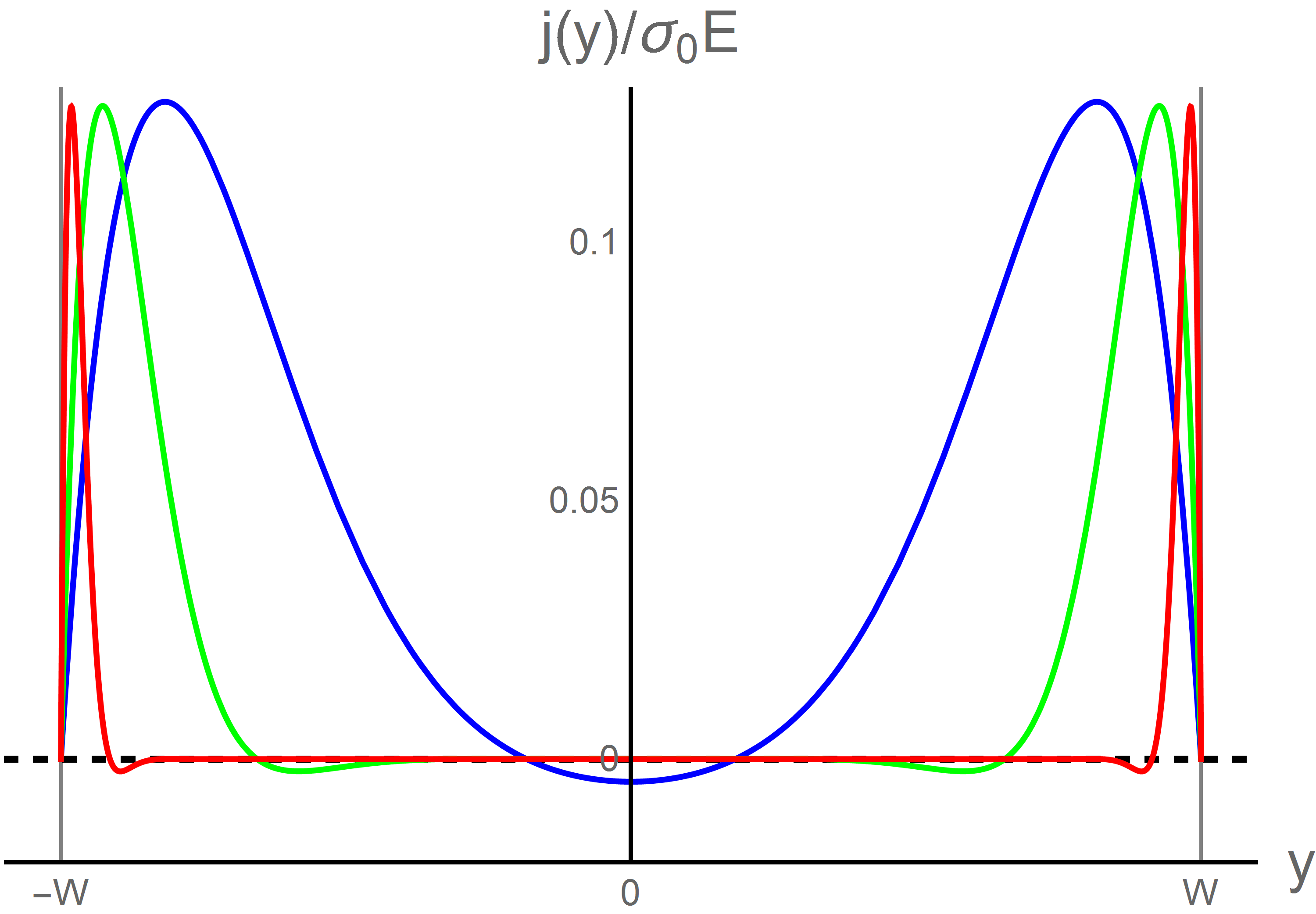}
\qquad\qquad\qquad
\includegraphics[width=0.4\textwidth]{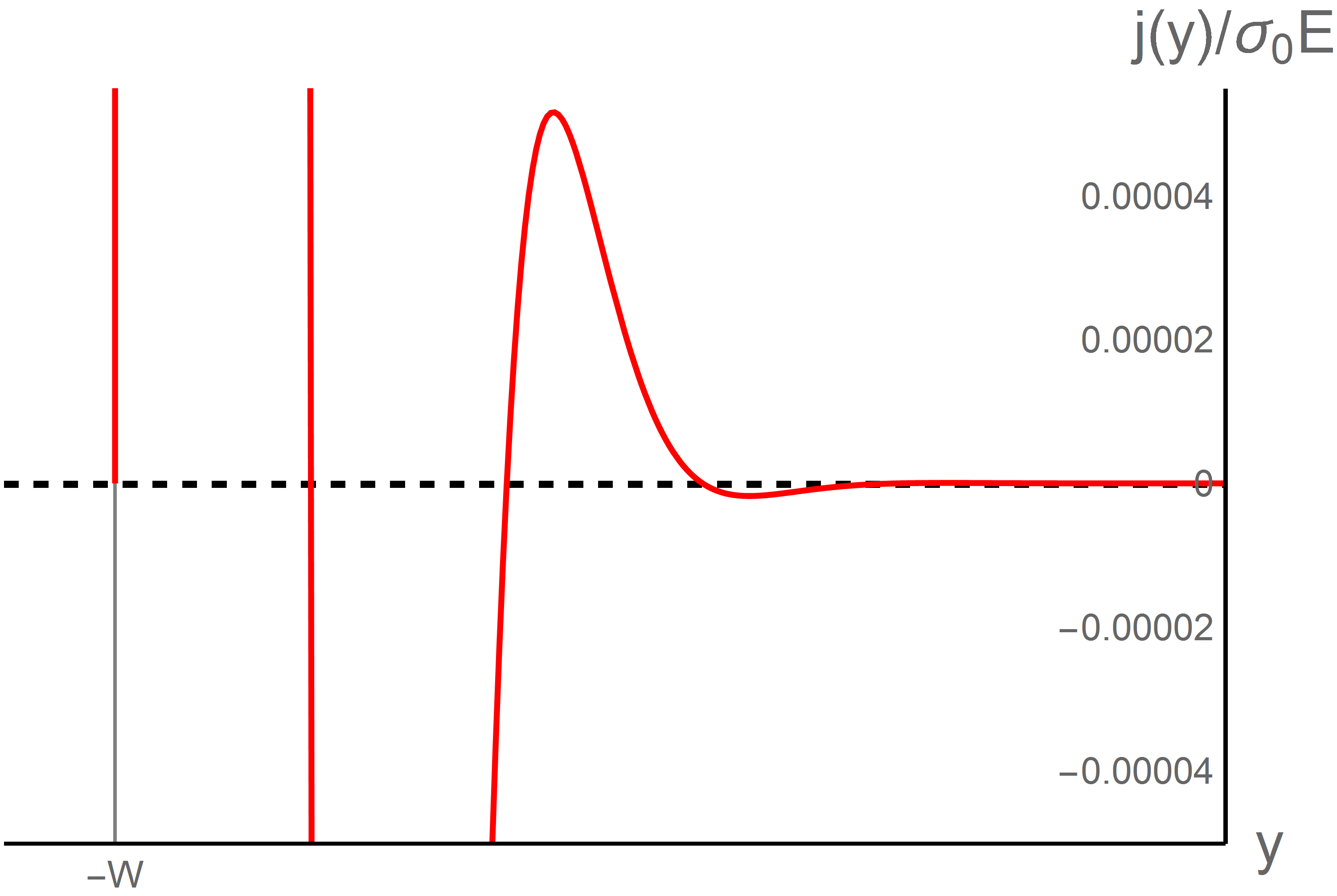}
}
\caption{Electric current density (\ref{rj2}) in the regime of fast
  recombination (\ref{cr}) and strong enough magnetic fields, $B>B_*$.
  Left panel: evolution of the current density with increasing
  magnetic field. The three curves (blue, green, and red) correspond
  to ${\omega_c\tau_{ee}\!=\!4,10,40}$, respectively, calculated with
  the values ${\ell_G(0)/\ell_R\!=\!0.25}$, ${W/\ell_R\!=\!0.25}$, and
  ${\tau\tau_*/\tau_{ee}^2\!=\!300}$.  Right panel: the fine structure
  of the edge current in strong magnetic field in the narrow range of
  values near zero. The numerical values correspond to the choice
  ${\ell_G(0)/\ell_R\!=\!0.25}$, ${W/\ell_R\!=\!0.25}$,
  ${\tau\tau_*/\tau_{ee}^2\!=\!1000}$, and
  ${\omega_c\tau_{ee}\!=\!50}$. The vertical grid lines indicate the
  sample edges and the horizontal dashed line indicates the zero
  value, ${j\!=\!0}$.}
\label{fig2:j}
\end{figure*}

In the absence of the magnetic field the equations (\ref{heqs4})
decouple and one finds the trivial solution [with the real
eigenvalues (\ref{lambda})]
\begin{equation}
\label{rj1}
j = \sigma_0 E \left[1-\frac{\cosh y/\ell_G(B)}{\cosh W/[\ell_G(B)]}\right],
\qquad
P=0,
\end{equation}
exhibiting the Poiseuille-like flow of the electric current.

\section{Counterflow of charge carriers in strong magnetic fields}

In the presence of the magnetic field, the eigenvalues (\ref{lambda})
may become complex. Indeed, using the relation
\[
\ell_G(0)\ll\ell_R,
\]
following from combining the assumption $\tau_{ee}\ll\tau_R$ [see
  Eq.~(\ref{cond})] and the fact that $\tau>\tau_*$ [by definition
  (\ref{s0})], we may re-write the eigenvalues (\ref{lambda}) as
\begin{equation}
\label{lapp}
\lambda_\pm \!\approx\!
\frac{1}{2\ell_G^{2}(B)}\left[1 \!\pm\!
\sqrt{\!1\!-\!4\ell_G^{2}(B)\ell_R^{-2}\omega_c^2\tau\tau_*}\right]\!.
\end{equation}
The behavior of the eigenvalues as functions of the magnetic field is
controlled by a parameter
\begin{equation}
\label{xi}
\xi=\frac{\ell_G^2(0)}{\ell_R^2}\frac{\tau\tau_*}{\tau_{ee}^2}\sim\frac{\tau_*^2}{\tau_R\tau_{ee}}.
\end{equation}
As long as $\xi<1$, the eigenvalues, $\lambda_\pm$, are real. In
Ref.~\onlinecite{usp}, we have assumed a stronger inequality,
$\xi\ll1$, and explored the resulting magnetoresistance.

\subsection{Oscillating currents}

Here we are interested in the regime of relatively fast recombination,
$\xi>1$. In this case, there exists a particular value of the magnetic
field, $B_*$, where the expression under the square root in
Eq.~(\ref{lapp}) vanishes:
\begin{equation}
  \label{bstar}
  \omega_c^* = \frac{1}{2\tau_{ee}\sqrt{\xi\!-\!1}}.
\end{equation}
For $B>B_*$, the eigenvalues (\ref{lapp}) are complex,
\begin{subequations}
\label{lpc}
\begin{equation}
\lambda_\pm(B\!>\!B_*)=\frac{1\pm i \gamma}{2\ell^{2}_G(B)},
\end{equation}
where
\begin{equation}
\label{gamma}
\gamma = \sqrt{\frac{B^2\!-\!B_*^2}{B_*^2}
\frac{1}{1\!+\!4\omega_c^2\tau_{ee}^2}}.
\end{equation}
\end{subequations}
As a result, the currents (\ref{rj2}) and (\ref{rp2}) acquire an
oscillating contribution.

Most interestingly, in the electric current (\ref{rj2}) the
oscillation amplitude may exceed the uniform background $j_0$ leading
to appearance of a counterflow, i.e. a {\it locally negative} current
density, see Fig.~\ref{fig2:j}. Here we plot the current density
(\ref{rj2}) for a case where the sample width is of the same order of
magnitude as the Gurzhi length, ${W\sim\ell_G(0)}$. In this case, the
counterflow first appears in the middle of the sample (see the blue
curve in the left panel Fig.~\ref{fig2:j}). As the field is increased,
the Gurzhi length $\ell_G(B)$ decreases and the current flow is being
pushed out towards the sample edges. If the width of the sample is
much larger than the Gurzhi length, ${W\gg\ell_G(0)}$, then the
current is flowing mostly along the edges \cite{usp} and therefore the
counterflow appears in the edge region.

Further analysis is greatly simplified for very strong fields, $B\gg
B_*$, and in the regime of fast recombination
\begin{equation}
\label{cr}
\xi\gg1
\quad\Leftrightarrow\quad
\tau_{ee} \ll \tau_R \ll \tau_*^2/\tau_{ee}.
\end{equation}
However, in this case there exists an intermediate field range
(absent for $\xi\gtrsim1$)
\begin{equation}
\label{intr}
\omega_c^*\ll\omega_c\ll\tau_{ee}^{-1},
\end{equation}
such that the imaginary part $\gamma$ exhibits two distinct types of
behavior
\[
\gamma \approx
\begin{cases}
B/B_*, & \omega_c^*\tau_{ee}\ll\omega_c\tau_{ee}\ll1, \cr
\sqrt{\xi}, & \omega_c\tau_{ee}\gg1.
\end{cases}
\]
In both cases the eigenvalues become purely imaginary
\begin{equation}
\label{lpm}
\lambda_\pm(B\gg B_*)
=
\pm \, i\ell^{-2}_c(B). 
\end{equation}
As a result, the spatial distribution of the currents is governed by
the single field-dependent length scale, $\ell_c(B)$:
\begin{eqnarray}
\label{lc}
&&
\ell_c^2(B) = \frac{\ell^2_G(B)}{\sqrt{\xi} }
\begin{cases}
(\omega_c\tau_{ee})^{-1}, & \omega_c\tau_{ee}\ll1, \cr
2 , & \omega_c\tau_{ee}\gg1,
\end{cases}
\\
&&
\nonumber\\
&&
\qquad\qquad
\sim \langle v^2\rangle \sqrt{\tau_R\tau_{ee}^3}
\begin{cases}
(\omega_c\tau_{ee})^{-1}, &  \omega_c\tau_{ee}\ll1, \cr
(\omega_c\tau_{ee})^{-2}, & \omega_c\tau_{ee}\gg1.
\end{cases}
\nonumber
\end{eqnarray}

Substituting the imaginary eigenvalues (\ref{lpm}) into the current
density (\ref{rj2}) we find the strongly damped oscillatory behavior
illustrated in Fig.~\ref{fig2:j}. When the characteristic scale of the
oscillations is much smaller than the width of the system, $\ell_c\ll
W$, we may expand Eq.~(\ref{rj2}) near the sample edges to find the
asymptotic expression
\begin{equation}
\label{je}
j(y) \approx \sigma_0 E\frac{\ell_c^2(B)}{\ell_G^2(B)} 
\sin\left(\frac{W/2 \!-\! |y|}{\sqrt{2}\ell_c}\right)
e^{-\frac{W/2-|y|}{\sqrt{2}\ell_c}}.
\end{equation}
At the same time, in the middle of the sample the current density is
equal to $j_0$ (up to exponentially small corrections). In strong
magnetic fields, $j_0$ is small (in absolutely pure samples it vanishes)
\begin{equation}
\label{jzero}
j_0(B) \approx \sigma_0 E/(\omega_c^2\tau\tau_*)\ll\sigma_0 E,
\qquad
j(\tau\!\rightarrow\!\infty)\!=\!0,
\end{equation}
such that the bulk current in strong enough magnetic field is almost
zero at the scale of the figure. This behavior is illustrated by the
red curve in  Fig.~\ref{fig2:j}.

The right panel in Fig.~\ref{fig2:j} illustrates the peculiar
structure of the edge current. While in the outermost region, the
current is positive, i.e. directed along the external electric field,
there is another, inner region carrying {\it negative} current flowing
in the direction {\it opposite} to the electric field. The existence
of this region stems from the oscillatory behavior (\ref{je}). These
oscillations, however, are strongly damped by the exponential decay
that occurs on the scale that is exactly the oscillation period (in
strong enough magnetic field). As a result, already the first minimum
of the expression (\ref{je}) is strongly suppressed leading to the
smallness of the negative current seen in Fig.~\ref{fig2:j}. In
principle, the bulk current is also oscillating, as illustrated in the
right panel in Fig.~\ref{fig2:j}.

\subsection{Counterflow threshold}

The eigenvalues (\ref{lambda}) remain real in zero field and become
complex only for ${B>B_*}$. Hence, the counterflow is a threshold
phenomenon. 

In weak fields the current density is positive everywhere in the
system. As the field is increased past $B_*$, the current density
develops oscillations. The magnitude of the oscillations grows with
the field and at some particular field, ${B_0>B_*}$, the current
density reaches zero at some point in the sample, ${j(y_0; B_0)=0}$,
such that at stronger fields the counterflow is developed around
$y_0$. 

For not too wide samples with ${W\sim\ell_G}$ the counterflow appears
around ${y_0=0}$, see Fig.~\ref{fig2:j}. For wider samples with
${W\gg\ell_G}$ the counterflow appears close to the edges,
${|y_0|\sim{W}/2}$. In the latter case, both $B_0$ and $y_0$ can be
found analytically in the limit (\ref{cr}). Substituting the
eigenvalues (\ref{lpc}) into the current density (\ref{rj2}), we find
near the edges, i.e. for $|y|\sim W/2$,
\[
j = \frac{j_0}{\gamma} {\rm Im}
\left[ \left( 1 \!-\! e^{\sqrt{1+i\gamma}\frac{|y|-W/2}{\sqrt{2}\ell_G(B)}} \right)
  \left(\frac{1}{2}\!+\!\omega_c^2\tau\tau_* \!+\! i \frac{\gamma}{2}\right)\right].
\]
This expression can be simplified as follows. For $\xi\gg1$, one finds
from Eq.~(\ref{bstar})
\[
\omega_c^*\tau_{ee}\ll1
\quad\Rightarrow\quad
\ell_G(B) \approx \ell_G(0),
\quad
\gamma \approx \sqrt{\frac{B^2}{B_*^2}\!-\!1}.
\]
Now, denoting
\[
\delta = \frac{|y|-W/2}{\sqrt{2}\ell_G(B)}, \quad
\sqrt{1+i\gamma}= c_1+ic_2,
\]
we re-write the current density in the form
\[
j = \frac{j_0}{\gamma} \left[
  \frac{\gamma}{2}\left(1 \!-\! e^{-c_1\delta}\cos c_2\delta\right)
  + \omega_c^2\tau\tau_* e^{-c_1\delta}\sin c_2\delta\right].
\]
Since the first term is less than unity, this expression first
vanishes at the point ${c_2\delta=3\pi/2}$. Substituting this into the
current, we find the equation for $\gamma$:
\[
\gamma = 2\omega_c^2\tau\tau_* e^{-\alpha},
\quad
\alpha = \frac{3\pi}{2}\frac{c_1}{c_2} = \frac{3\pi}{2} \frac{1\!+\!\sqrt{1\!+\!\gamma^2}}{\gamma},
\]
that can be re-written as an equation for ${x=B_0/B_*>1}$:
\[
\sqrt{x^2\!-\!1} = x^2 \frac{\ell_R^2}{2\ell_G^2(0)} e^{-\frac{3\pi}{2}\sqrt{\frac{x+1}{x-1}}}.
\]
Since ${\ell_R/\ell_G(0)\gg1}$, this equation does not admit large
solutions ${x\gg1}$. For ${x\!-\!1\ll1}$, the equation simplifies to
\[
a z = e^z, \quad a = \frac{\ell_R^2}{6\pi\ell_G^2(0)}, \quad
z = \frac{3\pi}{\sqrt{2(x-1)}}.
\]
This equation has two solutions for ${a>e}$, out of which we have to
choose the solution ${z>1}$ to be consistent with the assumption
${x\!-\!1\ll1}$. In general, the solutions of the above transcendental
equation cannot be expressed in terms of elementary functions. The
solution ${z>1}$ is given by the so-called Lambert $W$-function
\[
z = - W_{-1}(-1/a).
\]
For large $a$, we can use the asymptotic expression
\[
z \approx \ln a + \ln \ln a + {\cal O}(1),
\]
and as a result
\[
B_0 \approx B_* \left[ 1 + \frac{9\pi^2}{4(\ln a + \ln \ln a)^2}\right].
\]
In the extreme case where ${\ln a\gg1}$, the threshold field $B_0$ is
rather close to $B_*$. Otherwise, ${B_0/B_*\sim{\cal O}(1)}$.

\begin{widetext}

\subsection{Stability analysis}

The existence of regions with counterpropagating currents implies
the inhomogeneous distribution of the Joule heating across the
sample. Indeed, the work done by the external electric field is given
by the standard expression ${\bs{j}(y)\!\cdot\!\bs{E}}$, which is
becomes negative if the direction of the current flow is opposite to
that of the field. However, given the smallness of the negative
currents, see Eq.~(\ref{je}) and Fig.~\ref{fig2:j}, the overall work
of the external electric force is positive, ${IE>0}$. This can be seen
either by direct integration of the result (\ref{rj2}), or by using
the equations (\ref{heqs4}) to express the integrated inhomogeneous
current density in terms of the integral over the positive definite
quadratic form,
\begin{eqnarray}
\label{heat}
IE \!=\!\!\int\limits_{-W/2}^{W/2}\!\!dy j(y) E =
\frac{1}{\sigma_0}\!\!\int\limits_{-W/2}^{W/2}\!\!dy \left[j^2(y)
  \!+\! \ell_G^2(B) [j'(y)]^2\right]
+
\frac{\tau_*}{\sigma_0\tau}\!\!\int\limits_{-W/2}^{W/2}\!\!dy
\left[P^2(y) \!+\! \ell^2_R [P'(y)]^2\right],
\end{eqnarray}
which demonstrates the positivity of the work $IE$ irrespective to the
particular form of the solution $j(y)$. Hence, the system does not
develop any instability (in contrast to the case of the Ohmic regime
with negative conductivity \cite{zrs}). The fact that in some part of
the sample the local Joule heating appears to be negative means that
the heat is being redistributed between different parts of the
electronic system. This process is accompanied by a lateral energy
flow. The corresponding energy current can be determined by solving
the nonlinear hydrodynamic equations (taking into account the Joule's
heat). This calculation is beyond the scope of the present paper and
will be reported elsewhere.

Similar arguments can be used to establish the stability of the
solution (\ref{rj2}) while allowing for charge fluctuations. Following
the standard procedure for stability analysis \cite{dau6}, we
introduce plane wave solutions in the form
\[
{\cal O}(x,y;t) \rightarrow {\cal O}(x,y) e^{i\omega t},
\]
for fluctuations of all currents and densities in Eqs.~(\ref{ceq}),
(\ref{eq0}) including the fluctuation of the charge density
${\delta{n}=n-n^{(0)}}$ (${n=n_h-n_e}$). These fluctuations induce an
electric field (the Vlasov field \cite{dau10}). In the simplest case
of a gated 2D sample in the limit of strong screening \cite{us2}, the
induced field is proportional to the gradient of the charge density
fluctuation, $\bs{E}_V=-(4\pi ed/\epsilon)\bs{\nabla}\delta n$, where
$d$ is the distance to the gate and $\epsilon$ is the dielectric
constant. The dependence on $x$ is dictated by the geometry of the
problem and is given by $e^{ik_xx}$. The eigenmode frequency
$\omega=\omega_l(k_x)$ has to be determined by solving
Eqs.~(\ref{ceq}), (\ref{eq0}) and can in principle be
complex. Stability of a given solution is determined by the sign of
the imaginary part of the frequency with stable solutions
corresponding to ${\rm Im}\;\omega_l(k_x)\geqslant0$. Substituting
the above {\it Ansatz} into Eqs.~(\ref{ceq}), (\ref{eq0}), we follow
the same steps leading to Eqs.~(\ref{heqs}), (\ref{heqs4}). As a
result, we arrive at the following equations for the amplitudes of the
current densities:
\begin{subequations}
\label{st1}
\begin{equation}
\label{st1p}
i\omega \bs{P} = - \bs{P}/\tau + \omega_c\left[\bs{j}\times\bs{e}_z\right]
+\eta_{xx}\Delta\bs{P} + \frac{\langle v^2\rangle}{2} 
\frac{1}{i\omega+1/\tau_R}\bs{\nabla}(\bs{\nabla}\!\cdot\!\bs{P}),
\end{equation}
\begin{equation}
\label{st1j}
i\omega \bs{j} = - \bs{j}/\tau_* + \omega_c\left[\bs{P}\times\bs{e}_z\right]
+\eta_{xx}\Delta\bs{j} + 
\frac{s^2}{i\omega}\bs{\nabla}(\bs{\nabla}\!\cdot\!\bs{j}), 
\qquad
s^2 = \frac{\langle v^2\rangle}{2} + \frac{4\pi e^2}{m\epsilon}\rho^{(0)}d.
\end{equation}
\end{subequations} 
The quantities in Eqs.~(\ref{st1}) are fluctuations around the
time-independent (steady state) solution and hence the external
electric field does not enter Eqs.~(\ref{st1}). As a result, we have a
system of homogeneous linear equations which has nontrivial solutions
only if the determinant of the system is equal to zero. The latter
equation yields the allowed frequencies $\omega_l(k_x)$. Our goal,
however, is more modest -- we just need to establish the sign of
${{\rm{Im}}\;\omega_l(k_x)}$. To that end, we multiply
Eq.~(\ref{st1p}) by $\bs{P}^*$, take a complex conjugate of
Eq.~(\ref{st1j}) and multiply by $\bs{j}$, then add the resulting
equations and integrate over the area of the sample. As a result we
find the following relation:
\begin{eqnarray}
\label{st2}
&&
i\omega \!\int\!dxdy|\bs{P}|^2 -i\omega^*\!\int\!dxdy|\bs{j}|^2
=
-\frac{1}{\tau}\!\int\!dxdy|\bs{P}|^2-\frac{1}{\tau_*}\!\int\!dxdy|\bs{j}|^2
+\eta_{xx}\!\int\!dxdy\left(\bs{P}^*\Delta\bs{P}+\bs{j}\Delta\bs{j}^*\right)
\nonumber\\
&&
\nonumber\\
&&
\quad\qquad\qquad\qquad\qquad\qquad\qquad\qquad
+
\frac{\langle v^2\rangle}{2}\frac{1}{i\omega+1/\tau_R}
\!\int\!dxdy\bs{P}^*\!\cdot\!\bs{\nabla}(\bs{\nabla}\!\cdot\!\bs{P})
-
\frac{s^2}{i\omega}
\!\int\!dxdy\bs{j}\!\cdot\!\bs{\nabla}(\bs{\nabla}\!\cdot\!\bs{j}^*).
\end{eqnarray}
The last three terms can now be integrated by parts with the boundary
terms vanishing due to the no-slip boundary conditions, e.g.
\[
\int\!dxdy\bs{P}^*\!\cdot\!\bs{\nabla}(\bs{\nabla}\!\cdot\!\bs{P})
=
\int\!dxdy|\bs{\nabla}\!\cdot\!\bs{P}|^2.
\]
After that the only complex quantity in the equation is the
frequency. Introducing its real and imaginary parts,
${\omega=\omega_1+i\omega_2}$, we can separate the real part of the
equation and find the relation
\begin{eqnarray}
\label{st3}
&&
\omega_2\!\int\!dxdy\left[|\bs{P}|^2+|\bs{j}|^2
+
\frac{s^2}{\omega_1^2+\omega_2^2}|\bs{\nabla}\!\cdot\!\bs{j}|^2
+
\frac{\langle v^2\rangle}{2}
\frac{|\bs{\nabla}\!\cdot\!\bs{P}|^2}{\omega_1^2+(\omega_2-1/\tau_R)^2}
\right]
=
\frac{1}{\tau}\!\int\!dxdy|\bs{P}|^2+\frac{1}{\tau_*}\!\int\!dxdy|\bs{j}|^2
\nonumber\\
&&
\nonumber\\
&&
\quad\qquad\qquad\quad\qquad\qquad
+
\eta_{xx}\!\int\!dxdy \sum_{\alpha=x,y}\left(|\bs{\nabla}P_\alpha|^2+|\bs{\nabla}j_\alpha|^2\right)
+
\frac{\langle v^2\rangle}{2\tau_R}\frac{1}{\omega_1^2+(\omega_2-1/\tau_R)^2}
\int\!dxdy|\bs{\nabla}\!\cdot\!\bs{P}|^2.
\end{eqnarray}
Given that every single term in Eq.~(\ref{st3}) is manifestly
positive, we conclude that for every possible solution of the linear
system (\ref{st1})
\begin{equation}
\label{str}
\omega_2={\rm Im}\;\omega_l(k_x) >0,
\end{equation}
proving the stability of our theory. Note that this conclusion is
independent of the values of $\tau$ and $\tau_{eh}$ and remains valid
even in the limit $\tau,\tau_{eh}\rightarrow\infty$.

\section{Magnetoresistance}

Integrating the current density (\ref{rj2}) over $y$, we find the
total current, $I$, and hence the sample resistance \cite{usp}, $R$:
\begin{equation}
  \label{R}
  R \!=\! \frac{R_0 (\lambda_+\!-\!\lambda_-)}{
    \left[1\!-\!\frac{2\tanh\left(\sqrt{\lambda_+}W/2\right)}{W\sqrt{\lambda_+}}\right]
    \left[\ell_G^{-2}(B)\!-\!\frac{\lambda_-}{1+\omega_c^2\tau\tau_*}\right]
    \!-\!
    \left[1\!-\!\frac{2\tanh\left(\sqrt{\lambda_-}W/2\right)}{W\sqrt{\lambda_-}}\right]
    \left[\ell_G^{-2}(B)\!-\!\frac{\lambda_+}{1+\omega_c^2\tau\tau_*}\right]},
  \qquad
  R_0 \!=\! \frac{L}{e\sigma_0W}.
\end{equation}
\end{widetext}

The general expression (\ref{R}) for the sample resistance was
analyzed in Ref.~\onlinecite{usp} in the case of weak recombination
with the real eigenvalues (\ref{lambda}). Here we focus on the
opposite case of strong recombination, where $\lambda_\pm$ may take
the complex values (\ref{lpc}). The result is illustrated in
Figs.~\ref{fig3:r1} and \ref{fig4:r2}, where we show the dependence of
the sample resistance on the external magnetic field.

\subsection{Negative magnetoresistance}

\begin{figure}[t!]
\centerline{\includegraphics[width=0.9\columnwidth]{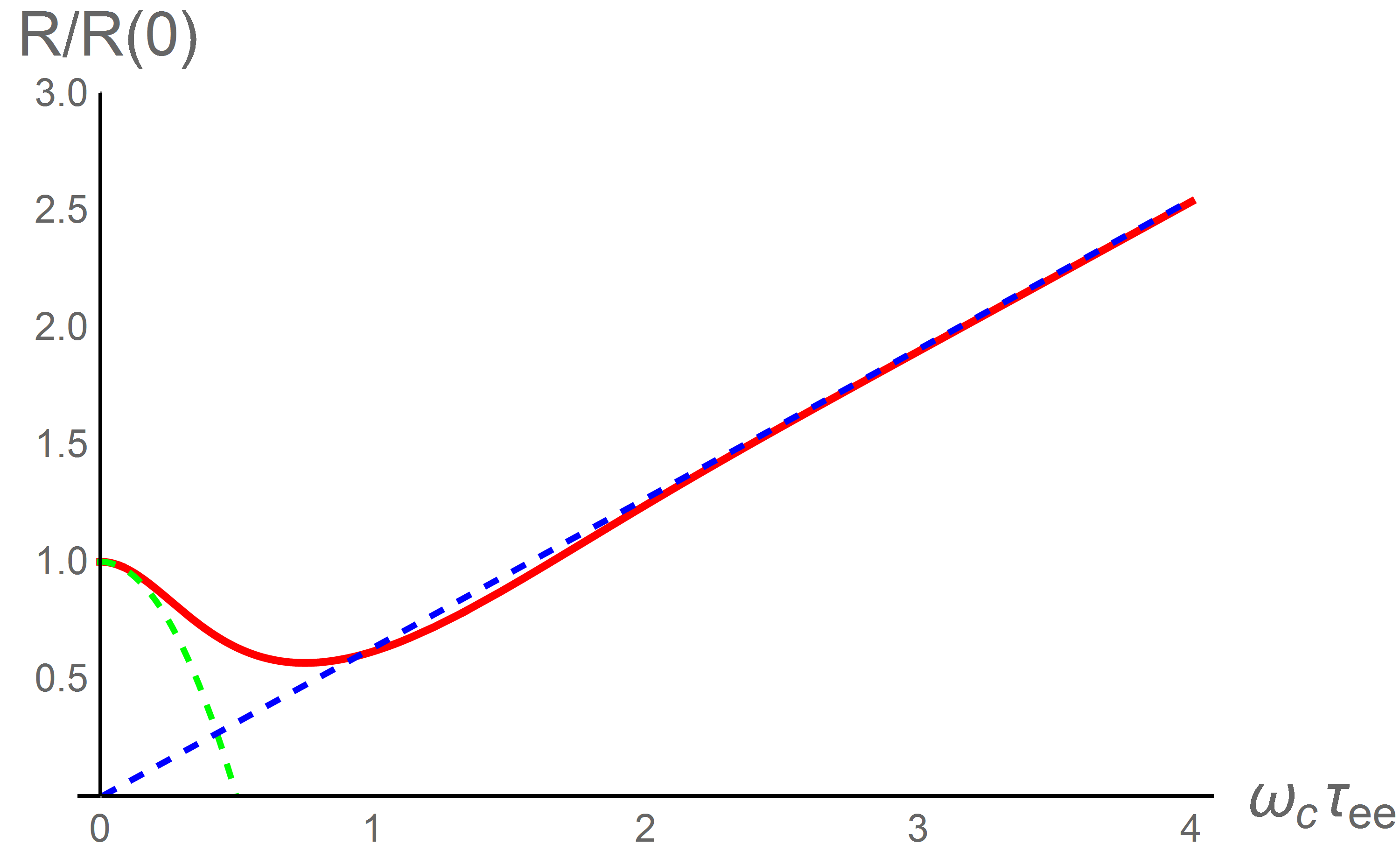}}
\caption{Sample resistance (\ref{R}) as a function of the magnetic
  field. The numerical values correspond to the following choice of
  parameters: ${\ell_G(0)/\ell_R\!=\!0.2}$, ${W/\ell_R\!=\!0.2}$,
  ${\tau\tau_*/\tau_{ee}^2\!=\!1000}$. The green dashed line indicates
  the negative, parabolic magnetoresistance in weak fields \cite{usp},
  see Eq.~(\ref{f3-p}). The blue dashed line shows the positive,
  linear magnetoresistance (\ref{r1}).}
\label{fig3:r1}
\end{figure}

In Fig.~\ref{fig3:r1} we illustrate the regime of negative
magnetoresistance similar to that discussed in
Ref.~\onlinecite{usp}. In weak fields, the eigenvalues (\ref{lambda})
remain real and the resistance (\ref{R}) decreases parabolically
\cite{usp}. For the choice of parameter values in Fig.~\ref{fig3:r1},
i.e. $W\leqslant\ell_G(0)$, the parabolic field dependence is given by
\cite{usp}
\begin{subequations}
  \label{f3}
  \begin{equation}
    \label{f3-p}
    R(B\rightarrow0)/R(0)=1-4\omega_c^2\tau_{ee}^2.
  \end{equation}
This behavior is shown in Fig.~\ref{fig3:r1} by the green dashed line.
  
In strong fields, the resistance grows linearly (i.e. the
magnetoresistance is positive). Similarly to the results of
Ref.~\onlinecite{usp}, we find (for ${\omega_c\tau_{ee}\gg{\rm max}[1,\ell_G(0)/(W\sqrt{\xi})]}$)
\begin{equation}
  \label{r1}
  R\approx R_0 A \left[\omega_c\tau_{ee} \!-\! A \frac{\tau_{ee}^2}{\tau\tau_*}\right],
\end{equation}
where
\begin{equation}
  A = -i\frac{W\gamma}{\ell_G(0)\sqrt{2}}
    \left[\frac{1}{\sqrt{1\!+\!i\gamma}}\!-\!\frac{1}{\sqrt{1\!-\!i\gamma}}\right]^{-1}.
\end{equation}
This behavior is shown in Fig.~\ref{fig3:r1} by the blue dashed line.
In the limit (\ref{cr}), i.e. for $\gamma\rightarrow\infty$, the
coefficient $A$ simplifies to
\begin{equation}
A(\gamma\rightarrow\infty) \approx \frac{W\xi^{3/4}}{2\ell_G(0)} .
\end{equation}
\end{subequations}

\subsection{Intermediate power-law regime}

\begin{figure}[t!]
\centerline{\includegraphics[width=0.9\columnwidth]{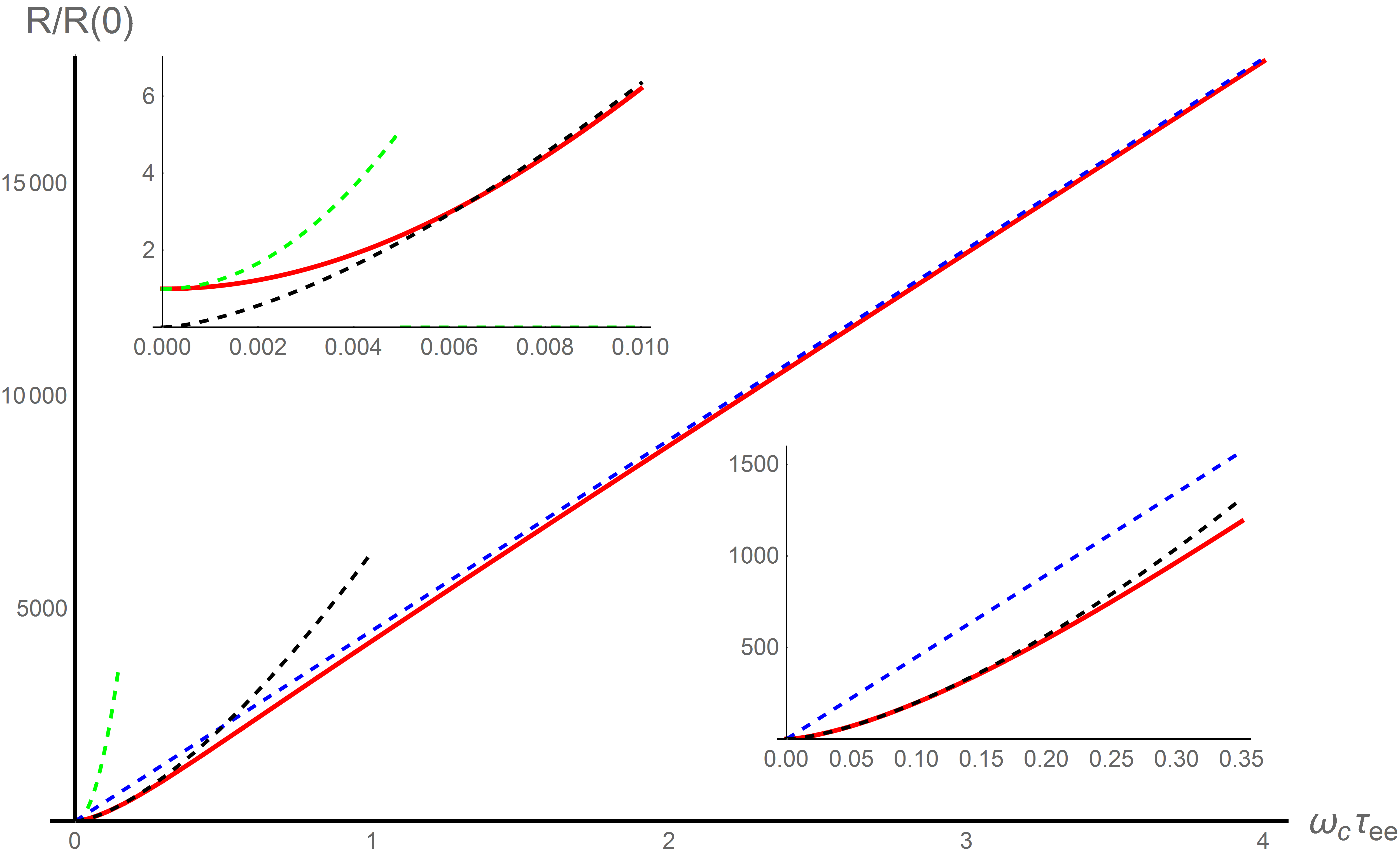}}
\caption{Sample resistance (\ref{R}) as a function of the magnetic
  field for ${\ell_G(0)/\ell_R\!=\!0.005}$, ${W/\ell_R\!=\!0.01}$,
  ${\tau\tau_*/\tau_{ee}^2\!=\!10^{10}}$. The dashed lines indicate
  the three asymptotic regimes: parabolic (green), Eq.~(\ref{f4-p});
  linear (blue), Eq.~(\ref{r1}); and the power law, $R\sim B^{3/2}$,
  (black) Eq.~(\ref{r32}). The insets zoom into the range of weak (up)
  and intermediate (down) fields.}
\label{fig4:r2}
\end{figure}

In Fig.~\ref{fig4:r2} we illustrate the regime of positive
magnetoresistance focusing on the limit (\ref{cr}). In this case, in
addition to the parabolic and linear asymptotics (in weak and strong
fields, respectively) discussed in Ref.~\onlinecite{usp}, we find an
additional regime appearing in the intermediate field range
(\ref{intr}).

For the parameter values used in Fig.~\ref{fig4:r2},
${\ell_R\gg{W}>W_0}$ (where
${W_0\approx[48\ell_R^2\ell_G(0)\tau_{ee}^2/(\tau\tau_*)]^{1/3}}$ is
the width where magnetoresistance changes sign), the parabolic
dependence of the resistance in the weakest fields is \cite{usp}
\begin{equation}
  \label{f4-p}
  \frac{R(B\rightarrow0)}{R(0)}=1+A_1\omega_c^2\tau_{ee}^2,
  \qquad
  A_1 = \frac{W^2}{12\ell_R^2}\frac{\tau\tau_*}{\tau_{ee}^2}.
\end{equation}
This behavior is shown in Fig.~\ref{fig4:r2} by the green dashed line
and is a good approximation only in the very weak fields, $B<B_*$, see
the upper inset. In stronger fields the eigenvalues (\ref{lambda})
become complex.

In the strongest fields, the resistance (\ref{R}) recovers the linear
behavior (\ref{r1}) shown in Fig.~\ref{fig4:r2} by the blue
dashed line.

In the intermediate field range (\ref{intr}), the eigenvalues
(\ref{lambda}) are linear in $B$, see Eq.~(\ref{lc}). Then, for wide
enough samples, ${W\gg\ell_c(B)}$, the field dependence of the
resistance (\ref{R}) is dominated by the power law with the exponent
$3/2$:
\begin{equation}
  \label{r32}
  \frac{R(B)}{R_0W}\approx \frac{\sqrt{\ell_G(0)}}{\sqrt{2}\ell_R^{3/2}} (\tau\tau_*)^{3/4}
    \omega_c^{3/2},
\end{equation}
shown in Fig.~\ref{fig4:r2} by the black dashed line. This behavior
appears only in the limit (\ref{cr}) of very strong recombination. For
weaker recombination, $\xi>1$, the field range (\ref{intr}) does not
exist, the eigenvalues (\ref{lambda}) do not develop the linear
behavior, and as a result the power law $R\sim B^{3/2}$ does not
appear.

\section{Qualitative discussion}

The nonuniform current distribution discussed in this paper bears a
certain similarity to the nonuniform spin density near a surface of a
three-dimensional semiconductor sample \cite{spin}. In that case, the
inhomogeneous spin density is created optically at the surface and
then propagates into the bulk of the sample by mans of carrier
diffusion. The equations of the spin diffusion in magnetic field
derived in Ref.~\onlinecite{spin} are similar to Eqs.~(\ref{heqs4}) if
one neglects electron-hole and disorder scattering. The resulting spin
density shows an inhomogeneity similar to that in Eq.~(\ref{je})
exhibiting oscillating behavior near the surface that decays into the
bulk of the sample.

\begin{figure*}[t!]
\centerline{\includegraphics[width=0.67\textwidth]{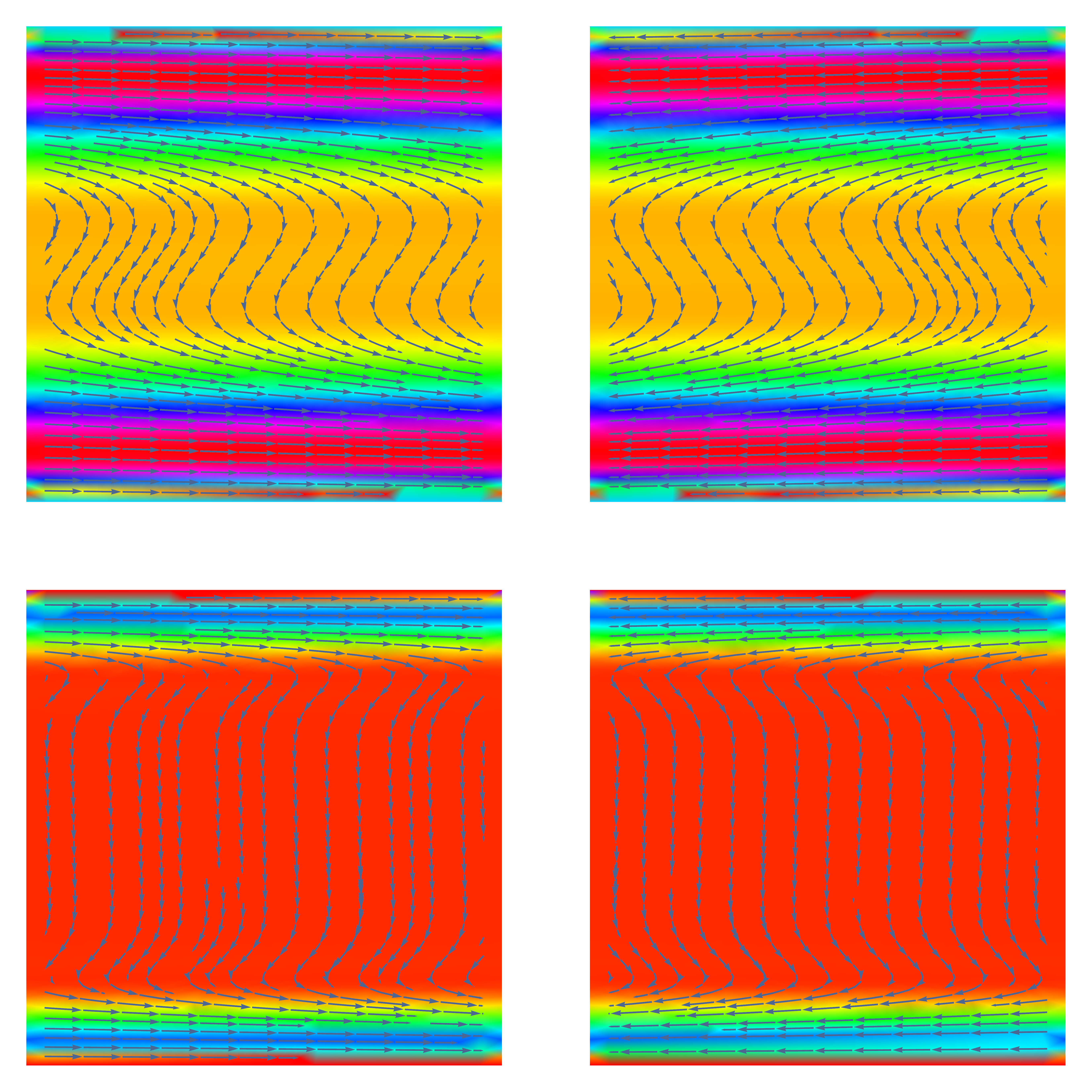}}
\caption{Individual electron (left) and hole (right) flows. The
  numerical values correspond to the same choice of parameters as in
  Fig.~\ref{fig2:j}: ${\ell_G(0)/\ell_R\!=\!0.25}$,
  ${W/\ell_R\!=\!0.25}$, ${\tau\tau_*/\tau_{ee}^2\!=\!300}$. The top
  panels show data for ${\omega_c\tau_{ee}\!=\!4}$ (same as the blue
  curve in the left panel in Fig~\ref{fig2:j}), the bottom panels show
  data for ${\omega_c\tau_{ee}\!=\!10}$ (same as the green curve in
  the left panel in Fig~\ref{fig2:j}). }
\label{fig5:maps}
\end{figure*}

Physically, the counterflow in compensated semimetals appears due to
the influence of the strong magnetic field on the motion of charge
carriers. A non-quantized magnetic field tends to bend semicalssical
trajectories away from the direction of the applied electric field. In
the context of an inviscid two-component system (e.g., a nearly
compensated semimetal) this effect was discussed in
Ref.~\onlinecite{us3}. Taking into account viscous effects, we find
that away from the boundary electrons and holes follow nontrivial
trajectories illustrated in Fig.~\ref{fig5:maps}.

In strong magnetic fields (bottom panels in Fig.~\ref{fig5:maps}),
both electrons and holes in the bulk of the sample are moving across
the sample such that the the combined electric current nearly
vanishes, see Eq.~(\ref{jzero}), while the quasiparticle current
$\bs{P}$ is nearly uniform. This is consistent with earlier
discussions of the effect of non-quantizing magnetic field on graphene
\cite{meg,usg} or compensated semimetals \cite{us2,us3,vas} and should
be contrasted with the usual interpretation of the classical Hall
effect in single-component electronic systems. In the latter case, the
lateral electric current which would be induced by the magnetic field
in an infinite system is compensated by the Hall voltage and as a
result, electrons move only in the direction of the applied electric
field. In a two-component system at charge neutrality the Hall voltage
is absent so that the electric currents in the two constituent
subsystems have to cancel each other.

In the edge region, the electron and hole currents experience a
rotation: very close to the edge [within the distance of the order of
  the Gurzhi length $\ell_G(B)$], the charge carriers move along the
boundary, while in the next layer (controlled by the quasiparticle
recombination) the lateral component of the currents appears. As a
result, the current vectors exhibit an intricate rotation pattern:
first they overshoot the angle $\pi/2$ between the edge and bulk
flows, but never reaching the angle $\pi$ and eventually return to
$\pi/2$. This relatively complex pattern is on one hand, required by
vanishing quasiparticle current $\bs{P}$ at the edge, but on the other
hand, appears due to the presence of the viscous layer. In an inviscid
system \cite{us2,us3}, there is no zero boundary condition on the
tangential component of the electric current and hence the electron
and hole currents rotate smoothly with their angle relative to the
boundary varying from $0$ to $\pi/2$.

In weaker fields, the above rotation pattern is incomplete due to the
overlap of the two boundary regions (i.e. ${\ell_c\sim{W}}$). In this
case the pronounced bulk region with the transverse moving charge
carriers does not develop, see the top panels in
Fig.~\ref{fig5:maps}. As a result, the counterflow occupies not the
edge, but a central region of the sample.

The nonuniform (and rotating) flows of electrons and holes are
characterized by the non-vanishing
$\bs{\nabla}\!\times\!\bs{j}_{e(h)}$. This should not be confused with
the true vorticity in the sense of a whirlpool (or eddy) formation
\cite{exg1,fal}. In fact, already the standard Poiseuille flow
\cite{dau6,poi} possess the nonvanishing $\bs{\nabla}\!\times\!\bs{v}$
(where $\bs{v}$ is the hydrodynamic velocity). However, the Poiseuille
flow is incompressible with $\bs{\nabla}\!\cdot\!\bs{v}\!=\!0$. As a
result, the transverse component of the velocity vanishes exactly,
$v_y=0$, and neither the true vorticity, nor any other rotation of the
velocity vector may appear. If the fluid exhibiting the Poiseuille
flow is charged (e.g., in a plasma with heavy ions that provide the
effective positive background to the electronic fluid), then the
vanishing of the transverse velocity component is ensured by the Hall
voltage. Now, in the two-component system with quasiparticle
recombination both the electron and hole fluids are compressible,
$\bs{\nabla}\!\cdot\!\bs{j}_{e(h)}\ne0$, with the total electric
current fulfilling $\bs{\nabla}\!\cdot\!\bs{j}=0$ (due to charge
conservation). In this case, the electron and hole currents exhibit
the rotation shown in Fig.~\ref{fig5:maps}, which is the ultimate
reason for the oscillating behavior (\ref{je}).

\section{Summary}

To summarize, we have considered the viscous electronic flow in
compensated semimetals with strong quasiparticle recombination and in
confined geometries. While the sample resistance is qualitatively
similar to the case of weak recombination considered in
Ref.~\onlinecite{usp}, see Fig.~\ref{fig3:r1}, the current density
profile shows a qualitatively different behavior. The current is
flowing mostly (with exponential accuracy) along the sample edges. At
each edge, the current flow is nonuniform and consists of two
counterpropagating stripes, see Figs.~\ref{fig1:j0} and
\ref{fig2:j}. Such counterflows are expected to be more general than
the particular problem considered in this paper and may appear even in
single-component electron fluids either in the {\it ac} transport
\cite{moe,aac} or in the thermoelectric flow \cite{usf}.

The appearance of the small local current density in the direction
opposite to that of the applied electric field does not affect the
global thermoelectric properties of the sample. However, this is a
direct indication of the inhomogeneous distribution of the Joule
heating across the sample accompanied by a lateral energy flow. As the
effect of Joule heating is beyond linear response, a proper theory of
the thermoelectric phenomena in semimetals requires a solution of the
nonlinear hydrodynamic equations. Such a theory will be reported
elsewhere.

\acknowledgments

We thank M.I. Dyakonov, E.I. Kiselev, A.D. Mirlin, D.G. Polyakov,
J. Schmalian, and M. Sch\"utt for fruitful discussions. This work was
supported by the FLAG-ERA JTC2017 Project GRANSPORT, the Dutch Science
Foundation NWO/FOM 13PR3118 (MT), and the Russian Foundation for Basic
Research Grant 17-02-00217 (VYK). PSA, APD, and IVG acknowledge the
support by the Russian Science Foundation of the analysis of viscous
effects in 2D electron systems that led to the discovery of the
counterflows (Grant 17-12-01182). MT acknowledges the support from
ITMO visiting professor fellowship program. BNN acknowledges the
support by the MEPhI Academic Excellence Project, Contract
No. 02.a03.21.0005.

\end{document}